\documentclass[aps,preprint]{revtex4}
\usepackage{graphicx,amssymb}
\begin{document}
\setcounter{page}{1}

\title{Flux-Pinning Behaviors and Mechanism  According to Dopant Level in  (Fe, Ti) Paticle-Doped MgB$_2$ Superconductor}

\author{H. B. \surname{Lee}}
\email{superpig@pusan.ac.kr}
\thanks{Fax: +82-51-513-7664}
\author{G. C. \surname{Kim}}
\author{Young Jin \surname{Shon}}
\author{Dongjin \surname{Kim}}
\author{Y. C. \surname{Kim}}
\affiliation{Department of Physics, Pusan National University, Busan 46241, Korea}
\begin{abstract}
We have studied flux-pinning effects  of MgB$_2$ superconductor by doping (Fe, Ti) particles of which radius is 163 nm on average. 5 wt.\% (Fe, Ti) doped  MgB$_2$ among the specimens showed the best field dependence of magnetization  and 25 wt.\% one did  the worst  at 5 K . The difference of field dependence of magnetization of the two increased as temperature increased. 
Here we show experimental results of (Fe, Ti) particle-doped MgB$_2$ according to dopant level and the causes of the behaviors.   Flux-pinning effect of volume defects-doped superconductor was modeled in ideal state. During the study, we had to divide M-H curve of volume defect-dominating superconductor as three discreet regions for analyzing flux pinning effects, 
 which  are diamagnetic increase region, $\Delta$H=$\Delta$B region, and diamagnetic decrease region.  
  As a result, flux-pinning effects of volume defects decreased as dopant level increased over the optimal dopant level, which was caused by decrease of flux-pinning limit of a volume defect. And similar behaviors are obtained as dopant level decreased below the optimal dopant level, which was caused by the decreased number of volume defects. Comparing the theory with experimental results, deviations  increased as dopant level increased over the optimal dopant level, whereas the two was well matched on less dopant level than the optimal dopant level. The behavior is considered to be caused by segregation of volume defects. On the other hand, the property of over-doped specimens dramatically decrease as temperature increases, which is caused by double decreases of flux-pinning limit of a volume defect and segregation  effect.
 \end{abstract}



\maketitle

\section{Introduction} 
We have studied the flux-pinning effects  of MgB$_2$ superconductor by doping (Fe, Ti) particles of  which radius  is  163 nm on average \cite{Lee2, Lee4, Lee5}.  Investigating  field dependences of magnetization (FDM) of the doped MgB$_2$ specimens, 5 wt.\% doped specimen showed the best FDM and FDMs of other doped specimen became  poorer as dopant level increases or decreases. 
The behavior means that there was the optimal dopant level for the best FDM when MgB$_2$ were doped with the (Fe, Ti) particles. 
 Although the behaviors might be interesting enough, the exact mechanism has not been revealed. 

On the other hand, there were several reports that diamagnetic property  and critical current (J$_c$) were changed according to dopant level \cite{Kia, Nishio, Martinez, Sudesh, Tang}.  However,  detailed mechanism and  the cause have been not shown. According to our studies, one thing to note is that the optimal concentration of defects for best performance depended on the state of pinning sites that defects produced \cite{ }. For example, flux-pinning effects of  defects depended on which types,  what sizes, and how many. 

\begin{figure}
\vspace{0cm}
\begin{center}
\includegraphics*[width=11cm]{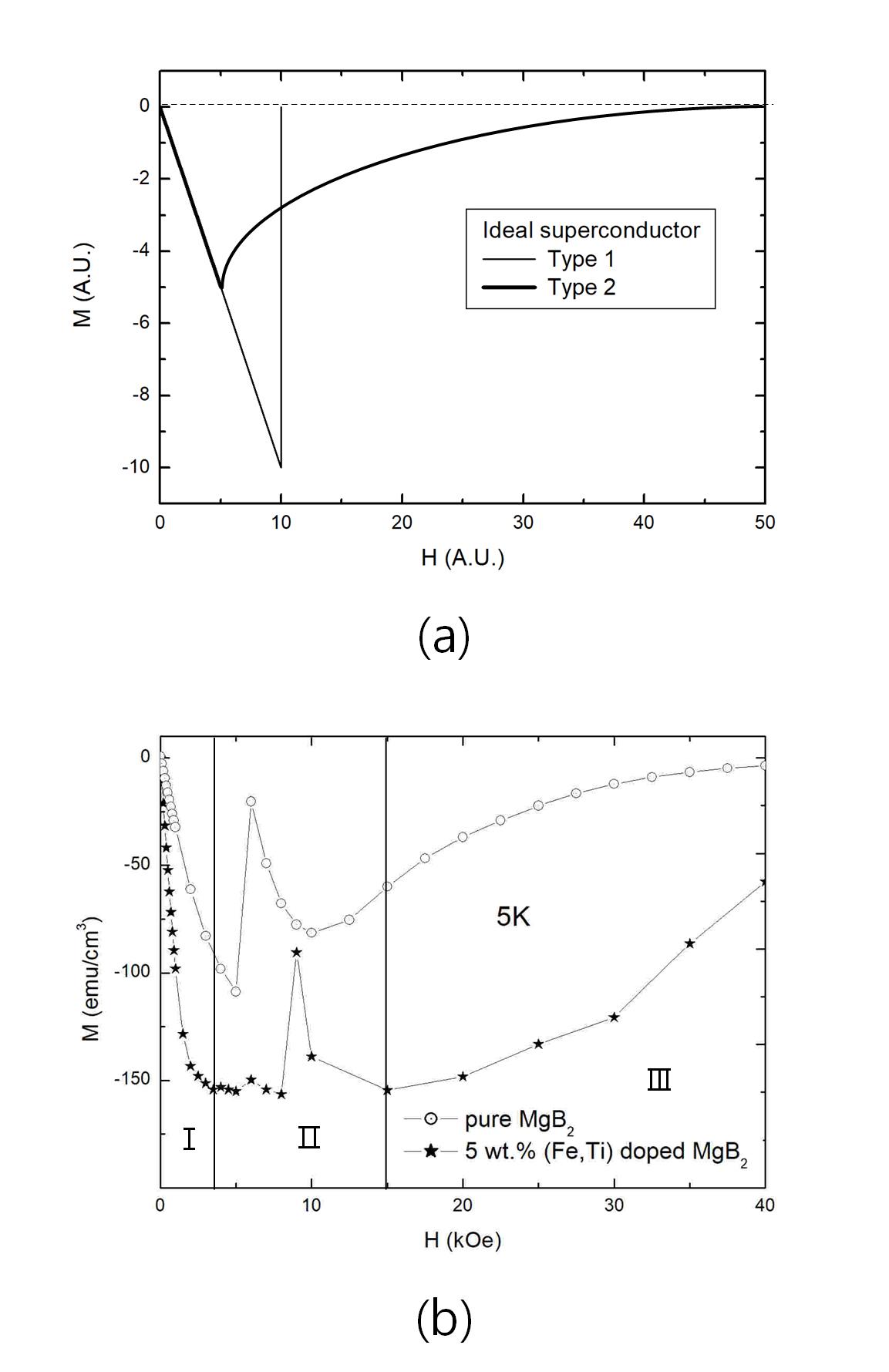}
\end{center}
\caption{ Field dependences of magnetizations (M-H curves) of superconductors (a): M-H curves of ideal type I and type II superconductors (b): M-H curve of 5 wt.\% (Fe, Ti) particle-doped MgB$_2$ at 5 K, which is volume defect-dominating superconductor. It can be divided as three discreet region, which is diamagnetic increase region (Region I),  $\Delta$H=$\Delta$B region (Region II), diamagnetic decrease region (Region III). M-H curve of pure MgB$_2$ at 5 K is used as a reference.}
\label{fig0}
\end{figure}

 The conventional theory for ideal type II superconductor represented two critical fields which are lower critical field (H$_{c1}$) and upper critical field (H$_{c2}$) in field dependence of magnetization curve (M-H curve) as shown Fig. 1 (a) \cite{Tinkham, Poole}. However,  M-H curves of current specimens, which are one of volume defect-dominating superconductors, showed quite different behavior from that of ideal superconductor \cite{ Nabialek, Khene, Liang, Chow, Radzyner, Daeumling}.  Thus, we have to explain  M-H curves  by dividing them into three discreet regions as shown Fig. \ref{fig0} (b). The first is the diamagnetic property increase region (region I), the second is  $\Delta$H=$\Delta$B region (region II), and the third is the diamagnetic  decrease region (region III). The cause of the distinction is that M-H curves of  real superconductors are heavily influenced by flux-pinning phenomena of defects and  each region has different mechanism  \cite{Lee4, Lee5}. 

Regarding the region of diamagnetic increase (region I), pinned fluxes at a volume defect are picked out (pick-out depinning) from the defect  when F$_{pickout}$ is larger that  F$_{pinning}$. The basis of diamagnetic increase after H$_{c1}$ is that the magnetic fluxes have penetrated into the superconductor after H$_{c1}$ are pinned at volume defects. Thus, they would be another barrier to prevent the fluxes penetrating into the superconductor if they are pinned at volume defects. %\cite{Lee5}. 

In  $\Delta$H=$\Delta$B region (region II),  a different cause is applied to the movement of pinned fluxes at volume defects.  The shortest distance between pinned fluxes at volume defects would determine whether the pinned fluxes have to be pick-out depinned or not. Thus they could be pick-out depinned  and move although F$_{pickout}$  is smaller than  F$_{pinning}$   \cite{Lee6}. Pick-out depinning was named by the behavior of depinning that pinned fluxes are picked out together when they are depinned from the volume defect.  The behavior of pinned fluxes in  $\Delta$H=$\Delta$B region are influenced by the nature that neighborhoods of a  volume defect would lose superconductivity if the shortest distance between pinned fluxes at the volume defect is the same as that of H$_{c2}$. Thus, flux-pinning itself is not established in the state and pinned fluxes at a volume defect are pick-out depinned although F$_{pickout}$  is smaller than  F$_{pinning}$. 

Regarding the diamagnetic decrease region (region III), it is the region that flux-pinning effect of volume defects decreases.
% 이영역이 단순히 Hc2 로 표현이 되는 지금의 표현방식 보다 중요한 이유는 같은 초전도체에서도 볼륨디펙트의 양과 상태에 따라 반자성 감소가 달라지기 때문이다. 
 In the region, magnetic fluxes which are not pinned at volume defects increase as applied magnetic field increases and the increase of unpinned fluxes results in a decrease of diamagnetic property of the superconductor by 4$\pi$M = B - H, where M, B and H are magnetization, magnetic induction, and applied magnetic field, respectively. The difference between conventional theory of ideal superconductor after H$_{c1}$ and region III of the current representation is the decrease rate of diamagnetic property.  The decrease rate was smaller as applied magnetic field increases when flux-pinning effects of volume defects increased (i.e. $\Delta$H=$\Delta$B region is wider).  

One may ask $''$It is natural that flux-pinning effects of volume defects decrease  if a dopant level increases because $\Delta$G$_{defect}$ is shallowed, which are caused by decreased portion of superconductivity.$''$ However,  the concept is considered to be came from the conventional view of superconductors. According to the concept of superconductor, $\Delta$G$_{defect}$ of 1 wt.\% doped MgB$_2$ and pure MgB$_2$ must have deeper $\Delta$G$_{defect}$ than those of  higher dopant level specimens. However, the results of current experiments are too far from it, which is 5 wt.\% doped MgB$_2$ showed deepest $\Delta$G$_{defect}$ in wider applied magnetic field. 
The basis of $\Delta$G$_{defect}$ is the field that represents the diamagnetic property ($\Delta$G$_{defect}$=$-\frac{(H-B)^2}{8\pi}$$\times$$\frac{4}{3}$$\pi$r$^3$, where $H$ is applied magnetic field and $B$ is magnetic induction).
Depending on kinds of  volume defects, $\Delta$G$_{defect}$ could be deeper if flux-pinning effect of the volume defects were good. And $\Delta$G$_{defect}$ could be shallower if flux-pinning effects of volume defects were poor. 
Therefore, it is not reasonable that the increased  number of volume defects caused $\Delta$G$_{defect}$ to be shallower.

In this paper, we would study the cause of the best field dependence of magnetization of 5 wt.\% (Fe, Ti) doped MgB$_2$ specimen and the cause of gradual decreases of field dependence of magnetization in other dopant level of  MgB$_2$ specimens as dopant level decreases or increases from the 5 wt.\%.

\section{results}
 \subsection{Experimental results of field dependences of magnetization for (Fe, Ti) particle-doped MgB$_2$ specimens}
\begin{figure}
\vspace{0cm}
\begin{center}
\includegraphics*[width=14cm]{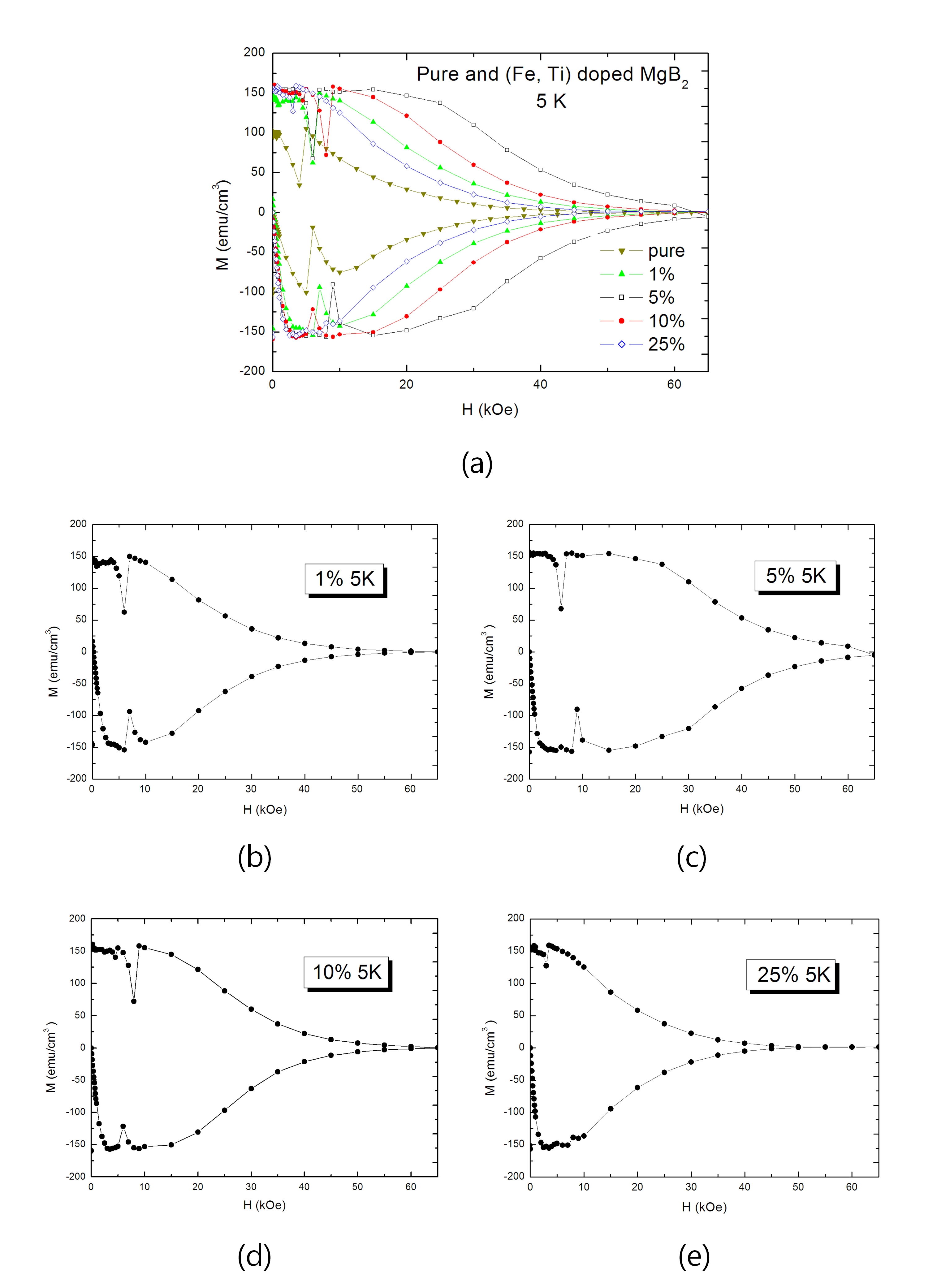}
\end{center}
\caption{ Field dependences of magnetizations (M-H curves) at 5 K of  wt.\% (Fe, Ti) particle-doped MgB$_2$ specimens, which was air-cooled. (a): M-H curves for comparison. (b): M-H curve of 1 wt.\% (Fe, Ti) particle-doped MgB$_2$. (c):  M-H curve of 5 wt.\% (Fe, Ti) particle-doped MgB$_2$. (d):  M-H curve of 10 wt.\% (Fe, Ti) particle-doped MgB$_2$. (e):  M-H curve of 25 wt.\% (Fe, Ti) particle-doped MgB$_2$. }
\label{fig1}
\end{figure}

Figure \ref{fig1} (a) shows M-H curves of (Fe, Ti) particle-doped MgB$_2$ specimens at 5 K according to dopant level, which are doped with (Fe, Ti) particles of which radius are  163 nm on average. Inspecting region I,  maximum diamagnetic properties (MDP) of the specimens except pure MgB$_2$ are almost same. However, they show different widths of region II, and 5 wt.\% doped specimens showed the widest $\Delta$H=$\Delta$B region. A width of the region is ordered as follows, 5\%$>$10\%$>$1\%$>$25\%$>$pure. Pure MgB$_2$ has no $\Delta$H=$\Delta$B region. 

Inspecting  the behaviors after $\Delta$H=$\Delta$B region (region III) in M-H curves as shown in the figure, the decrease rate of diamagnetic property along applied magnetic field was lower  as the width of the region was wider.
After 5 wt.\% doped specimen showed the lowest decrease rate of  field dependence of magnetization (FDM) along applied magnetic field,  decrease rates of other doped specimens increase as  dopant level increases from 5 wt.\%. And less doped specimens than 5 wt.\% also showed same behaviors as dopant level decreases from 5 wt.\%. 

\begin{figure}
\vspace{2cm}
\begin{center}
\includegraphics*[width=17cm]{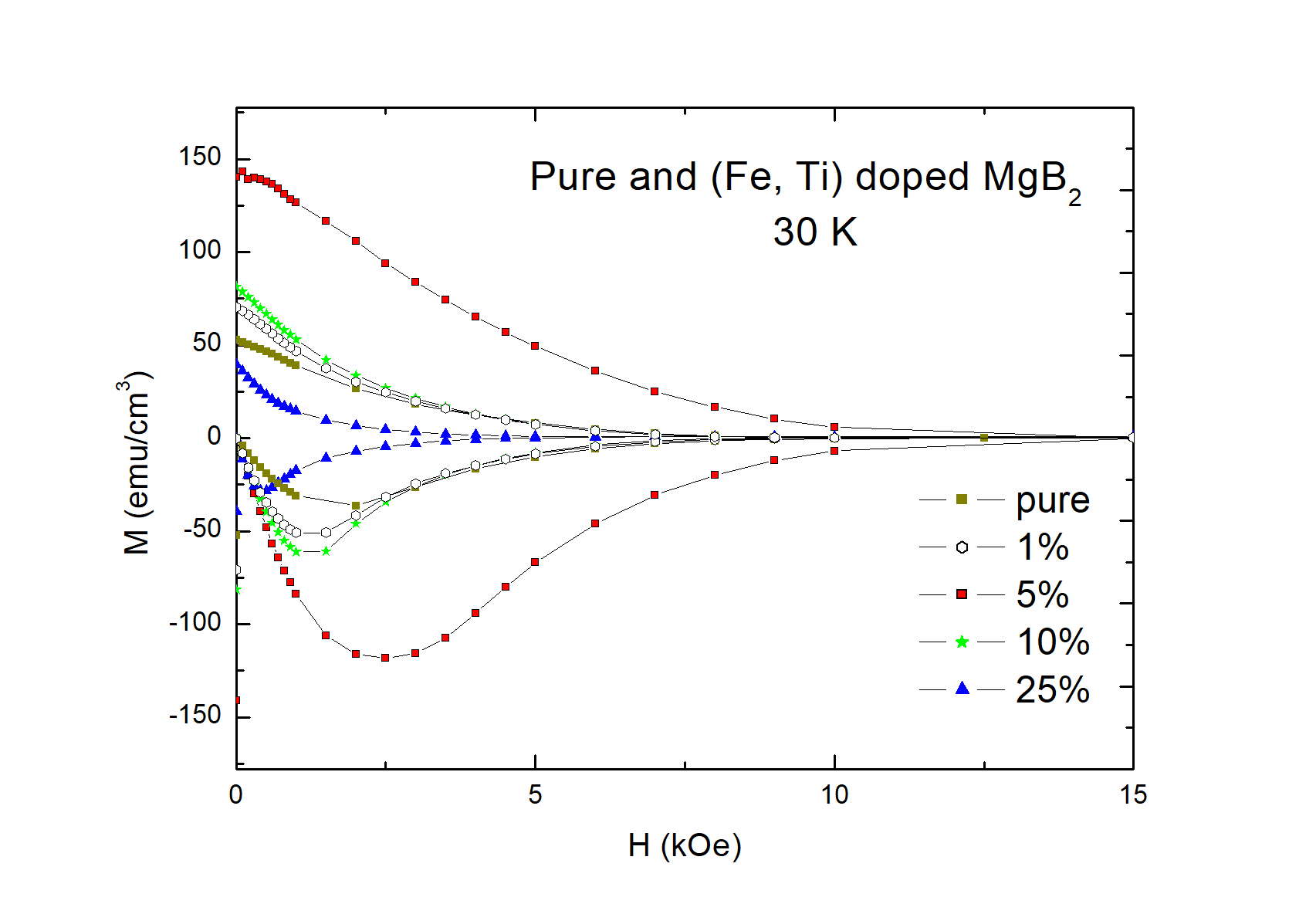}
\end{center}
\caption{ Field dependence of magnetizations (M-H curves) of  wt.\% (Fe, Ti) particle-doped MgB$_2$ at 30 K, which was air-cooled.}
\label{fig2}
\end{figure}

M-H curves of  doped MgB$_2$ specimens at 30 K are shown in Fig. \ref{fig2}, which is results of the same specimens as shown in Fig. \ref{fig1}. At the temperature, flux-pinning effects of a superconductor have to be determined by heights of the maximum diamagnetic property (MDP) and the degree of diamagnetic decrease because $\Delta$H=$\Delta$B region completely disappeared at 30 K. 
 The specimen showing the best maximum diamagnetic property at 30 K is still 5 wt.\% specimen, which had widest $\Delta$H=$\Delta$B region at 5 K. 
 However,  other specimens have some changes on M-H curves. The MDP of 10 wt.\% doped specimen is the next, which is same order when compared with a width of $\Delta$H=$\Delta$B region at 5 K, but  it is much closer to that of 1 wt.\% doped specimen. 

The figure  also shows that a difference of diamagnetic decrease between 5 wt.\% doped specimen and other specimens  at 30 K much more increased when compared with those at 5 K.  A noting thing is that 25 wt.\% doped specimen showed much poorer diamagnetic decrease at 30 K compared with that of pure MgB$_2$, which is reversed results at 5 K. From the considerations, it is understood that a degree of diamagnetic decrease at 30 K increases as dopant level increases from 5 wt.\% doped specimen. 

Considering the experimental results, it is needed to understand why 5 wt.\% specimen at 5 K showed the optimal flux-pinning effects, why gradual decrease of a width of $\Delta$H=$\Delta$B region occurred as the amount of dopant level increases or decreases from optimal dopant level,  why  a difference of diamagnetic decrease between the optimal flux-pinning specimen and other doped specimens  significantly increase when the temperature increases at 30 K, and why superconducting properties of doped specimens  go much poorer at 30 K as dopant level increases.

 \subsection{ A representation of flux-pinning effect in ideal doped state}
%\subsection{ A representation of flux-pinning effect in ideal doped state}

\begin{figure}
\vspace{2cm}
\begin{center}
\includegraphics*[width=8cm]{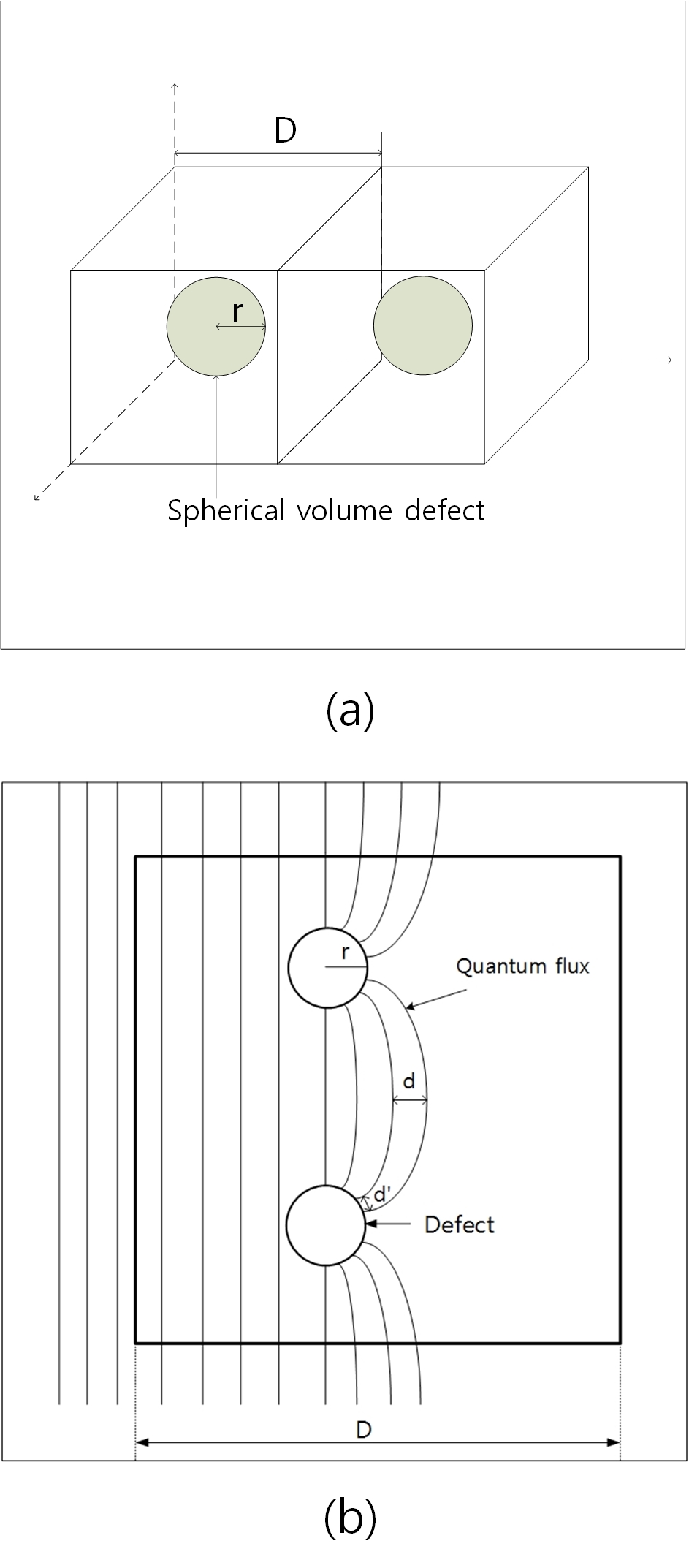}
\end{center}
\caption{(a): Schematic representation of doped volume defects in MgB$_2$ base. (b): Schematic representation of pinned fluxes at volume defects.}
\label{fig5}
\end{figure}

Figure \ref{fig5} (a) shows a schematic representation of volume defect-doped superconductor. 
It is assumed that spherical volume defects of which radius is $r$ are doped in cubic superconductor of which a length is $D$. Figure \ref{fig5} (b) shows a schematic representation which is several quantum fluxes  that are pinned together at volume defects. The shortest distance between pinned fluxes is  $d'$ and the widest one is  $d$. Generally, the number of quantum fluxes (n$^2$) which can be pinned at a spherical volume defect are calculated as follows. 
\begin{eqnarray}
n^2 = \frac{\pi r^2}{d'^2}
 \end{eqnarray} 
 where $r$ and $d'$ are radius of volume defect and the shortest distance between quantum fluxes pinned at the volume defect of which radius is $r$  when they have square structure, respectively  \cite{Lee5}.
If $d'^2$ is 2$\pi$$\xi^2$ ($\xi$ is coherence length of the superconductor), a volume defect pin the maximum flux quanta, which is flux-pinning limit of a volume defect. 

On the other hand, average $d$, which  is average value of widest distance between pinned fluxes  is 
\begin{eqnarray}
\bar{d} = \frac{D-r}{n}
 \end{eqnarray}
where $d$, $D$, $r$, and $n$ are the widest distance between pinned fluxes, a distance between volume defects,  radius of volume defect, and  the number of quantum fluxes  pinned at  volume defects in one dimension (1 D) in the state that they are arrayed in a row to the next volume defect as shown in Fig. \ref{fig5} (b).

 On the other hand, concerning the widest distance between pinned fluxes ($d$), minimum distance  must be needed for flux-pinning effect of volume defects, and the reasons are as follows.
 The first is that forefront fluxes among pinned fluxes have more tension, thus the foremost $d$ among pinned fluxes at volume defects (Fig.  \ref{fig5} (b)) must be shorter than average $d$. And the second is that pinned fluxes are affected by depinnings of other part of fluxes  from neighborhood  volume defects because pinned fluxes are interconnected from a volume defect to the next. Thus, pinned fluxes are vibrating if they are not depinned when other part of fluxes are depinned.  The third is that they are affected by heat caused by depinning of other fluxes and  other part of fluxes. Therefore, they cannot maintain flux-pinning state if the distance between pinned fluxes are shorter than the minimum distance, which must be much wider than that of H$_{c2}$.

The quantum flux pinned at  volume defects would be divided into two parts, which are volume defect part and superconducting part as shown in Fig. \ref{fig5} (b). Since quantum fluxes of volume defect part are fixed on the volume defects, they can withstand until the distance between them is same as that of H$_{c2}$  because $F_{pinning}$ (the pinning force) of 163 nm radius volume defect is much larger than $F_{pickout}$ (the pick-out depinning force) in $\Delta$H=$\Delta$B region \cite{Lee6}. However, quantum fluxes of superconducting part are less arrested on the volume defects because they are elastic and away from the volume defects. Therefore, they are highly affected by other fluxes movements and movements of other parts of quantum fluxes because a quantum flux is simultaneously pinned at 8000 volume defects in 5 wt.\% doped MgB$_2$, which result in  that  superconductivity of the part would disappear if the distance between the flux quanta  were shorter than that of H$_{c2}$ at any moment by the effects of environment. 

Quantum state of magnetic fluxes would be maintained by the repulsive force generated by superconducting eddy currents, and pinned fluxes at volume defect are constantly influenced by the movement of other parts of pinned fluxes \cite{Tinkham}.
If quantum fluxes failed to be separated by the influence of other fluxes,  the superconductivity of the part would disappear at the moment that the distance between fluxes is shorter than that of H$_{c2}$. The electrons that generate the magnetic flux would be not superconducting electrons anymore, which are Cooper pair. In order for normal electrons to generate  magnetic fluxes, it is certain that many electrons have to participate. 

Therefore, the heat  caused by electric resistance would happen. If the generated heat  was  small, it does not affect neighborhoods  of quantum fluxes to combine, thus they would retain their superconductivity again. However, if it was large enough, it would affect the neighborhoods of quantum fluxes to combine because coherence length of a superconductor increases as temperature increases. %온도의 상승은 코랭을 증가시키기 때문에 .
Thus,  the heat would propagate around it and causes other quantum fluxes to combine into one, which means that they are not a quantum flux anymore. At last, the entire  pinned fluxes at volume defect become non-superconducting state.  Thus, flux-pinning effects would disappear, and the pinned fluxes naturally are pick-out depinned from the defect. Therefore, some distance ($d$) between pinned fluxes at volume defects must be required  to maintain superconductivity  of pinned fluxes at a volume defect.

Therefore,  flux-pinning limit of a volume defects at over-dopant level is 
\begin{eqnarray}
n_{ov}^2 = n_o^2(\frac{D_{ov} - r}{D_o - r})^2
 \end{eqnarray}
where $n_o^2$,  $D_o$, and  $D_{ov}$ are the number of pinned fluxes on a volume defect at optimal dopant level at 0 K, the distance between volume defects at optimal dopant level, and the distance between volume defects at over-dopant level, respectively. It was assumed that pinned fluxes at a volume are arrayed vertically equal.

And a rate of increased pinning sites by over-doping is 
\begin{eqnarray}
R = \frac{m_{ov}^2}{m_o^2}
 \end{eqnarray}
where $m_o^3$ is the number of volume defects at optimal dopant level and $m_{ov}^3$ is  the number of volume defects at over-dopant level.
 The equation  was described because single flux quantum is pinned on  volume defects 
 along an axis. 
Therefore, total flux-pinning effects of volume defects at over-dopant level, which is expressed as a width of $\Delta$H=$\Delta$B region,  is
\begin{eqnarray}
W_{\Delta H= \Delta B, ov}=\frac{n_{ov}^2}{n_o^2}\frac{m_{ov}^2}{m_o^2} W_{\Delta H= \Delta B, o} = (\frac{D_{ov} - r}{D_o - r})^2 \frac{m_{ov}^2}{m_o^2}W_{\Delta H= \Delta B, o}
 \end{eqnarray}
where $W_{\Delta H= \Delta B, ov}$ and $W_{\Delta H= \Delta B, o}$ are a width of $\Delta$H=$\Delta$B region of over-dopant level and that of optimum level at 0 K, respectively.

On the other hand, under the state that volume defects are in less-dopant level, flux-pinning effects of the superconductor  depend on the number of volume defects because  flux-pinning limit of  a volume defect is the same as that of optimum level.
\begin{eqnarray}
W_{\Delta H= \Delta B, le}=\frac{m_{le}^2}{m_o^2}W_{\Delta H= \Delta B, o}
 \end{eqnarray}
where $W_{\Delta H= \Delta B, le}$ is a width of $\Delta$H=$\Delta$B region of less-dopant level and $m_{le}^3$ is the number of volume defects at less-dopant level in a superconductor.

Numerically,  5 wt.\% (Fe, Ti) particles in MgB$_2$ approximately corresponds to 2.0 vol.\% and $D$ is 5.94$r$.  On the other hand, 25 wt.\% (Fe, Ti) particles in MgB$_2$ corresponds to approximately 10 vol.\% and $D$ is 3.47$r$. A spherical volume defect of 163 nm radius in 5 wt.\% (Fe, Ti) particles-doped  MgB$_2$  can pin approximately 51$^2$ flux quanta when H$_{c2}$ is 65.4 Tesla (T) at 0 K \cite{Lee5}. Assuming that 51 quantum fluxes are arrayed in a row to the next volume defect, 
 $\bar{d}_{5 wt.\%}$ is 15.8 nm.

Table I shows various properties of (Fe, Ti)-doped  MgB$_2$ specimens according to dopant level.   $\bar{d}_{25 wt.\%}$ is 7.89 nm for 25 wt.\% (Fe, Ti)-doped  MgB$_2$ if 51$^2$ flux quanta are pinned at 163 nm radius volume defect. Superconducting state would be destroyed if $d$ is less than 5.63 nm  when H$_{c2}$ is 65.4 T at 0 K. 
Since the front $d$ in  pinned fluxes would be shorter than  $\bar{d}$ and the environments would affect $d$, it is certain that 163 nm radius volume defect cannot pin 51$^2$ flux quanta in dynamic state.

 5 wt.\% doped specimen  showed the best flux-pinning effect, and of which  $\bar{d}_{5 wt.\%}$  is 15.8 nm as mentioned. The number of pinned fluxes ($n^2$) at 163 nm radius defect would be approximately 26$^2$ in 25 wt.\% doped specimen if  $\bar{d}_{5 wt.\%}$  is set as the minimal distance. 
 The calculation means that 163 nm radius volume defect of 25 wt.\% doped specimen can pin up to only 26$^2$ flux quanta. Therefore, it is considered that the flux-pinning effect of a volume defect in 25 wt.\% doped specimen decreases to 0.26  (26$^2$/51$^2$) by Eq. (3). 
 
 On the other hand, increased pinning sites  of 25 wt.\% doped specimen is 2.92 (13680$^2$/8000$^2$) by Eq. (4) because single flux quantum is pinned on 13680 volume defects per unit length (cm) along an axis. Thus, actual rate of flux-pinning effect by increased volume defects for 25 wt.\% doped specimen are 0.75 (0.26$\times$2.92) by Eq. (5). Table I also shows actual flux-pinning effect of increased volume defects for 10 wt.\% doped specimen, which is 0.91. %(3.71x163/17.9=33.8, 34$^2$/45$^2$ =0.57, 10080$^2$/8000$^2$ =1.59, thus, 0.57x1.59 = 0.91). 
On the other hand, decreased flux-pinning effect of 1 wt.\% specimen is  0.37 (4860$^2$/8000$^2$) by Eq. (6). %For  1 wt.\% specimen, the calculation is possible because d of pinned fluxes at volume defects is rather wider than that of 5 wt.\% specimen. In other words, that is because the number of fluxes pinned at volume defects is same, but the number of volume defects decrease.

Comparing theoretical values with the experimental results as shown in Table I,   $\Delta$H=$\Delta$B region widths of 5 wt.\%, 10 wt.\%, 25 wt.\%, and 1 wt.\% are  experimentally 1.5 Tesla (T), 1.0 T,  0.5 T, and 0.6 T, respectively, as shown Fig. \ref{fig2} \cite{Lee7}. 
 1 wt.\% specimen  shows  a good match with theoretical value (0.6 T/1.5 T = 0.4). However,  as the dopant level increases from 5 wt.\%, a difference between the experimental results and the calculation increases (10 wt.\% specimen: 0.91 -  0.67 = 0.23, 25 wt.\% specimen: 0.75 - 0.33 = 0.42). This means that there would be other factors decreasing flux-pinning effect in over-doped specimens.

\subsection{ The segregation effect of volume defects  and temperature dependence of a width of $\Delta$H=$\Delta$B region at over-dopant level }
As the number of doped volume defects increases in a superconductor,  
average distance between them do not only decrease, but  volume defects which are closer than average distance also increase. %Thus, the concentration of pinned fluxes of the volume defects in the segregated area   would be high. When another quantum fluxes approach to the volume defect which has already  pinned fluxes, the fluxes would be pick-out depinned one by one and move into an inside of the superconductor. 
As discussed earlier, the shorter the distance between volume defects is, the fewer the pinning limit of a volume defect  is. Furthermore, if the distance between them is shorter than
$\sqrt{2\pi}$$\xi$  of the superconductor, they are recognized as a pinning site.  Thus, the volume defects look like being connected each other in the state. Therefore, the fluxes move more easier along the volume defects. %One noting thing is that the segregated volume defects  increases the L, which also affects the pinning limits of other volume defects.
%그래서 이들은 쭉 연결된 것 같은 느낌을 받을 것이고, 그래서 플럭스들은 더 쉽게 이동을 한다. 
 Consequently,  the fluxes can quickly penetrate into an inside of the superconductor when applied field increases because the segregated volume defects have lower pinning limits. %
 
The flux-pinning limit of a volume defect decreases as temperature increases, which is caused by increase of coherence length as temperature increases.
Therefore, a width of $\Delta$H=$\Delta$B region of optimal dopant level at a temperature  is
\begin{eqnarray}
W_{\Delta H= \Delta B(T),o}
%= \frac{n^2_{T}}{n_o^2}\frac{m_{o}^2}{m_o^2}W_{\Delta H= \Delta B(0), o}
= \frac{n^2_{T}}{n_o^2}W_{\Delta H= \Delta B(0), o}
\end{eqnarray}
where $n^2_T$ is flux-pinning limit of a volume defect at a temperature and $W_{\Delta H= \Delta B, o}$ is a width of $\Delta$H=$\Delta$B region of opimual level at 0 K. 

In the state of over-dopant level, a width of $\Delta$H=$\Delta$B region of  at a temperature is
\begin{eqnarray}
W_{\Delta H= \Delta B(T), ov}=%\frac{n_{ov(T)}^2}{n_{ov(0)}^2}\frac{m_{ov}^2}{m_{ov}^2}  W_{\Delta H= \Delta B(0), {ov}}=
\frac{n_{ov(T)}^2}{n_{ov(0)}^2} W_{\Delta H= \Delta B(0), {ov}}
 \end{eqnarray}

On the other hand, as temperature increases, over-doped superconductor would have much lower flux-pinning effect because volume defects have two kinds of decreases of flux-pinning limits. 
One  is a decrease of flux-pinning limit of a volume defect by over-doping , and the other is that of temperature increase, the two both are caused by increase of coherence length as temperature increases. Therefore, 
\begin{eqnarray}
W_{\Delta H= \Delta B(T), ov}=\frac{n_{ov(T)}^2}{n_{ov(0)}^2} W_{\Delta H= \Delta B(0), {ov}}
=  \frac{n_{ov(T)}^2}{n_{ov(0)}^2}\frac{n_{ov}^2}{n_o^2}\frac{m_{ov}^2}{m_o^2}W_{\Delta H=\Delta B(0), o}
 \end{eqnarray}
where $n^2_{ov}$ is flux-pinning limit of a volume defect of over-dopant level at 0 K. The last Equation came from Eq. (5). Therefore, flux-pinning effects ($W_{\Delta H= \Delta B(T), ov}$) superconductor at over-dopant level decrease dramatically as temperature increases compared with those of opitmal dopant level.

It is expected that M-H curves of over-doped specimens would be much poorer than those of optimal and less-doped specimen as temperature increases.  Figure \ref{fig2} shows those behaviors, which are that  the segregation effects support  that  M-H curve of the 25 wt.\% doped specimen showed  poorer M-H curve than that of pure specimen at 30 K although the M-H curve of 25 wt.\% doped specimen showed  better M-H curve than that of pure specimen at 5 K. 

\subsection{Discussion}

We suggested that M-H curves of real superconductors have to be considered as three discreet regions  owing to their different causes for the regions as shown Fig. \ref{fig1} (b). %They are different from that of  two critical field of ideal superconductor that is based on H$_{c1}$ and H$_{c2}$.
 If there were no defects in a superconductor,  which means no flux-pinning effect, the M-H curve  would reduce to ideal superconductor by no increasing diamagnetic property in Region I and deleting $\Delta$H=$\Delta$B region, which is Region II. And M-H curve become three regions if volume defects are many enough as being explained.   When planar defects are dominant in a superconductor, presence of Region II  depends on the dominance of planar defects. Region II completely disappears when dominance of planar defects are high, such as HTSC bulks which were made by solid state reaction method  because flux quanta move fast along grain boundaries although they have flux-pinning effects \cite{Senoussi, Seyoum}.%특히 Seyoum의 논문을 보면, 이 초전도체 엠에이치 커브는 저온에서의  엠지비투의 그것과 비슷하다. 이는 코랭의 증가로 인하여 피닝효과는 적고 침투효과는 크진 결과라고 이야기 할 수 있다cite{ }. 

However, Region II partially appears in the state that dominance of planar defects is low such as water-quenched MgB$_2$ specimen \cite{Lee8}. 
In addition, volume defects do not only increase diamagnetic property of the superconductor, but also form Region II. The behavior means that H$_{c2}$ state partially appears in Region II  by pinned fluxes at volume defects. The behavior is based on the fact that pinned fluxes at volume defects are pick-out depinned from the defect when the shortest distance between pinned fluxes is the same as that at H$_{c2}$ \cite{Lee4}. 

 Region III is the region that flux-pinning effects of the volume defects are weakened as mentioned.  Figure 5 (a) and (b) shows examples of the behavior.  They show diamagnetic properties of 5 wt.\%  and 10 wt.\% (Fe, Ti)-doped MgB$_2$  along applied magnetic field at various temperatures, respectively.  Generally, the better flux-pinning state of the volume defects are,  the wider  $\Delta$H=$\Delta$B region is formed in Region II \cite{Lee4}. Thus, the specimens have shown that wider $\Delta$H=$\Delta$B region induced  smaller decreases of diamagnetic properties in Region III as applied magnetic field increases. Therefore, it is determined that Region III also affected by flux-pinning effects of volume defects.

 The represented theory is only applicable to volume defects-dominating superconductor. All of bulk superconductors except single crystal have grains, which results in making grain boundaries. 
The air-cooled specimens of current experiment have also grains, thus they have grain boundaries of MgB$_2$. Nevertheless, the flux-pinning effects of grain boundaries are ignored at low field, which is less than 2 T.   It was because effect of volume defects are not only dominant at the field but also the width of grain boundary was short and amount of grain boundaries are small due to large grain size \cite{Lee8}.

\section{Conclusion}
We studied flux-pinning effects of  (Fe, Ti) particle-doped MgB$_2$ specimens according to dopant levels of (Fe, Ti) particles, of which radius is 163 nm on average.   M-H  curves of the specimens are able to be explained as three discreet regions, which are diamagnetic increase region after H$_{c1}$ (Region I), $\Delta$H=$\Delta$B region (Region II), and diamagnetic decrease region (Region III).  Flux-pinning effect of volume defects-doped superconductor was modeled in an ideal state that the amount of doped volume defects are varied from the optimal dopant level. We focused a width of $\Delta$H=$\Delta$B region  in comparing theoretical values with experimental results because  a degree of  diamagnetic decrease is inversely proportional to  a width of $\Delta$H=$\Delta$B region and max-diamagnetic properties of the doped specimens were almost same.

Results of experiments represented that 5 wt.\% (Fe, Ti)-doped MgB$_2$ showed widest $\Delta$H=$\Delta$B region and smallest diamagnetic decrease in diamagnetic decrease region among doped specimens, and widths of $\Delta$H=$\Delta$B region gradually became shorter as dopant level increases or decreases from 5 wt.\%. 
In addition, widths of $\Delta$H=$\Delta$B region became much shorter as temperature increases as dopant level increased. Comparing experimental results with theory, the two well matched at less-dopant level than optimal dopant level, but the difference between the two increased as dopant level increased from optimal dopant level.  Inspecting the cause  that  over-doping  of (Fe, Ti) particles would reduce flux-pinning effects, it was revealed that flux-pinning limit of a volume defect decreases as dopant level increases, and the segregation effect. %Pinned fluxes at volume defects could be divided as a part of volume defect and a part of superconductivity, and the latter  needs some distance between them. 
% 레스 도핑에 대해서는 어디에 집어 넣나? 

%The cause of the behavior is  that there is the minimal distance between pinned fluxes for flux-pinning effects in the ideal state, and the pinned fluxes are apt to be pick-out depinned from the defect  if the distance decreases below minimal level, which result in the decrease of flux-pinning limit of a volume defect. However, we could not explain all decreases of flux-pinning effect of over-doped specimen by the minimal distance. Another cause of the decrease  was determined to be a segregation of volume defect. If dopant level increases, the distance between volume defects which is shorter than average one would increases, and the short distance between defects was determined to be another cause of decrease of flux-pinning effects.

\section{Method}
 (Fe, Ti) particle-doped MgB$_{2}$ specimens were synthesized using the non-special atmosphere synthesis (NAS) method \cite{Lee1}. 
 The starting materials were Mg (99.9\% powder) and B (96.6\% amorphous powder) and (Fe, Ti) particles. Mixed Mg and B stoichiometry, and  (Fe, Ti) particles were added by weight. They were finely ground and pressed into 10 mm diameter pellets. (Fe, Ti) particles were ball-milled for several days, and average radius of (Fe, Ti) particles was about 0.163 $\mu$m. 
   On the other hand, an 8 m-long stainless- steel (304) tube was cut into 10 cm pieces. One side of the 10 cm-long tube was forged and welded. The pellets and excess Mg were placed in the stainless-steel tube. The pellets were annealed at 300 $^o$C  for 1 hour to make them hard before inserting them into the stainless-steel tube. The other side of the stainless-steel tube was also forged. High-purity Ar gas was put into the stainless-steel tube, and which was then welded. Specimens had been synthesized at 920 $^o$C  for 1 hour and cooled in air. Field dependences of magnetization were measured using a MPMS-7 (Quantum Design). During the measurement, sweeping rates of all specimens were made equal for the same flux-penetrating condition. 

\vspace{2cm}

%Acknowledgements\\  

\begin{figure}
\vspace{0cm}
\begin{center}
\includegraphics*[width=11cm]{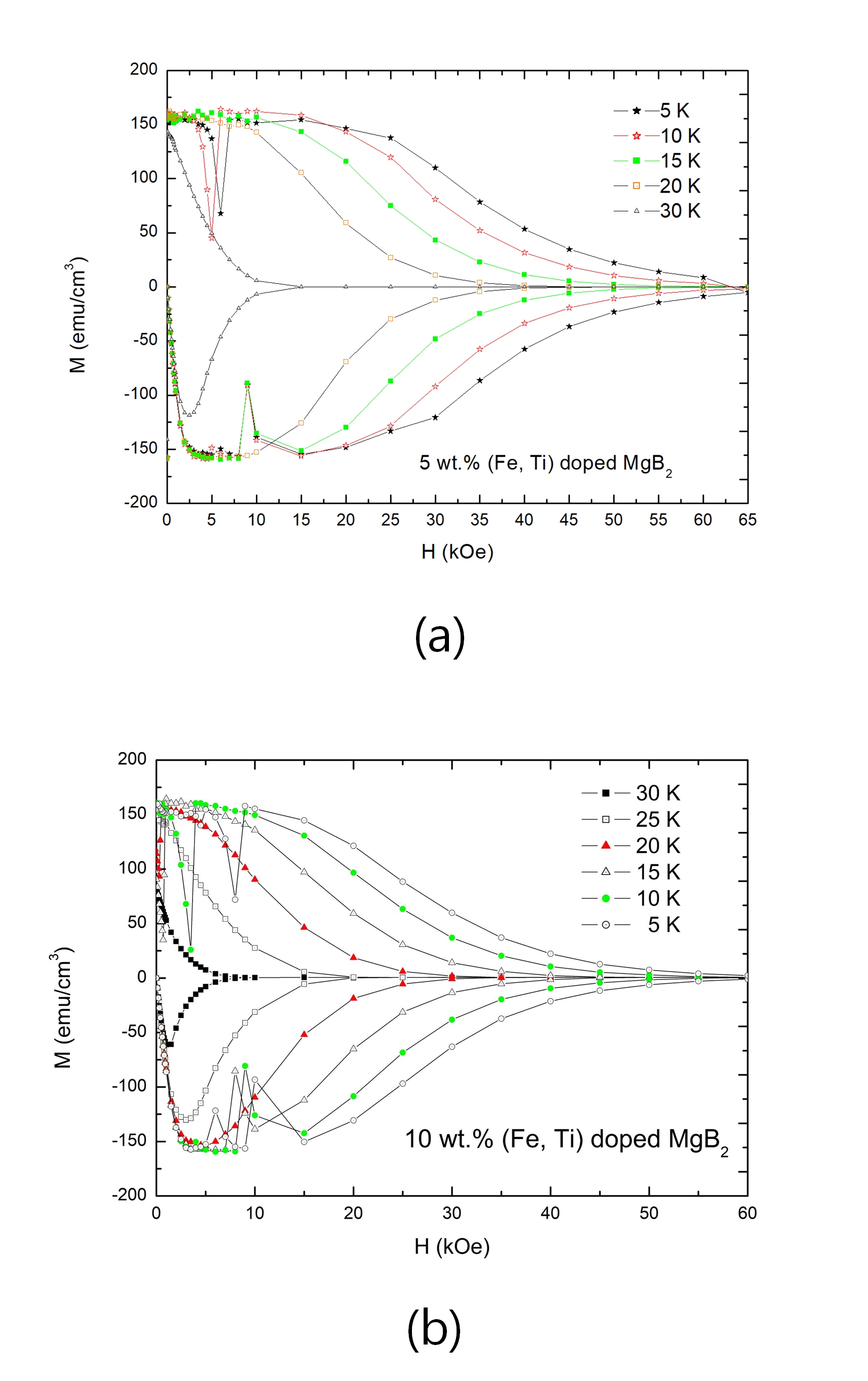}
\end{center}
\caption{(a): M-H curves of 5 wt.\% (Fe, Ti) particle-doped MgB$_2$ with variation of temperature. (b):  M-H curves of 10 wt.\% (Fe, Ti) particle-doped MgB$_2$ with variation of temperature. }
\label{fig4}
\end{figure}

\newpage
\begin{table}[!h]
\vspace{1cm}
\caption{ Various properties of (Fe, Ti) doped MgB$_2$ superconductors according to weight percentage of (Fe, Ti) particles, of which average radius is 163 nm. L$'$ and L  are the distance between defects in ideal state and closed packed state, respectively.  Closed packed state of volume defects is the state that all of fluxes penetrated are pinned at volume defects ($2r\times m_{cps}$=1, where $m_{cps}$ is the number of volume defects in close packed state \cite{Lee4}. Widths of $\Delta$H=$\Delta$B region of specimens are compared with that of 5 wt.\% (Fe, Ti) doped MgB$_2$ as a unit.}%폭은 5%를 기준으로 계산되었다. 
\begin{center}
\renewcommand{\tabcolsep}{18pt}
\renewcommand{\arraystretch}{0.5}
\begin{tabular}
{|c||c|c|c|c|}
%{|p{5cm}|p{3cm}|p{3cm}|p{3cm}|}\hline
\hline\hline
 & & & & \\
 Weight percentage &1 &$5 $& 10&25 \\
  & & & & \\
\hline\hline
 & & & & \\
 Volume percentage &0.4  &$2$& 4&10 \\
  & & & & \\
 
The number of  
& 4680$^3$& 8000$^3$& 10080$^3$& 13680$^3$\\
volume defects in cm$^3$& & & & \\
 & & & & \\
  The distance (L$'$, r=0.163 $\mu$m)  &10.15r &5.94r& 4.71r &3.47r\\ 
between volume defect & & & & \\
(Ideal state, $\mu$m) &1.65 &0.97 &0.68 & 0.57\\
  & & & & \\
  The distance (L) between & & & & \\
volume defect&14.0&$4.79$&3.02 &1.64 \\
 (Closed packed state, $\mu$m)   & & & & \\
& & & & \\
  Calculated flux-pinning effects& 0.37&1 &0.91 &0.75 \\
(Width of $\Delta$H=$\Delta$B region) & & & & \\
   & & & & \\
Experimental results& 0.4&1 &0.67 &0.33 \\
(Width of $\Delta$H=$\Delta$B region)& & & & \\
   & & & & \\
Differences between the calculated & 0.03&0 &0.24&0.42 \\
and  experimental results & & & & \\
   & & & & \\
Flux-pinning limit &51$^2$ &51$^2$ &38$^2$ &26$^2$ \\
of a volume defect& & & & \\
   & & & & \\
 \hline
\end{tabular}
\end{center}
\end{table}

\end{document}